# AI in Food Marketing from Personalized Recommendations to Predictive Analytics: Comparing Traditional Advertising Techniques with AI-Driven Strategies

Elham Khamoushi

*Department of English, Iowa state university, Iowa, USA*
*Corresponding author: elhamkh@iastate.edu*

**Abstract**

Artificial Intelligence (AI) has revolutionized food marketing by providing advanced techniques for personalized recommendations, consumer behavior prediction, and campaign optimization. This paper explores the shift from traditional advertising methods, such as TV, radio, and print, to AI-driven strategies. Traditional approaches were successful in building brand awareness but lacked the level of personalization that modern consumers demand. AI leverages data from consumer purchase histories, browsing behaviors, and social media activity to create highly tailored marketing campaigns. These strategies allow for more accurate product recommendations, prediction of consumer needs, and ultimately improve customer satisfaction and user experience. AI enhances marketing efforts by automating labor-intensive processes, leading to greater efficiency and cost savings. It also enables the continuous adaptation of marketing messages, ensuring they remain relevant and engaging over time. While AI presents significant benefits in terms of personalization and efficiency, it also comes with challenges, particularly the substantial investment required for technology and skilled expertise. This paper compares the strengths and weaknesses of traditional and AI-driven food marketing techniques, offering valuable insights into how marketers can leverage AI to create more effective and targeted marketing strategies in the evolving digital landscape.

## 1- Introduction

Artificial intelligence (AI) technology has revolutionized the food marketing landscape in recent years. Traditional advertising techniques are being complemented and, in some cases, replaced by artificial intelligence-driven strategies in food marketing. Marketing messages are being crafted, targeted, and delivered to consumers in a fundamentally different way as a result of this evolution. Artificial intelligence offers unprecedented capabilities for personalizing recommendations, predicting consumer behavior, and optimizing marketing efforts, such as machine learning, natural language processing, and predictive analytics. Food marketing traditionally relied heavily on mass media channels such as television, radio, print, and outdoor advertising[1]. The traditional methods of building brand awareness and reaching a wide audience have proven effective. One-size-fits-all approaches, however, can lead to inefficiencies and a lack of personalized engagement with consumers. Marketers are now able to create targeted, relevant, and engaging marketing campaigns thanks to the emergence of digital media and the subsequent rise of artificial intelligence technology[2]. Customized recommendation systems are one of the most prominent applications of artificial intelligence in food marketing[3]To deliver highly tailored product recommendations, these systems use data from multiple sources, including purchase history, browsing behavior, and social media interactions. AI-driven recommendation systems can analyze individual preferences and predict future behaviors more accurately than traditional advertising techniques, which rely on demographic data[4]. Increased customer satisfaction, enhanced consumer experience, and higher conversion rates are the benefits of this level of personalization. AI is also making a significant impact on predictive analytics. Marketers can anticipate consumer needs and behaviors using predictive analytics by using historical data to forecast future outcomes[5]. Predictive analytics can be used to identify trends in food marketing, optimize inventory management, and design more effective marketing campaigns. An AI algorithm can, for instance, predict which products will be popular at different points of the year based on seasonal trends, market demand, and consumer preferences. Marketers can use this information to tailor their campaigns accordingly, ensuring the right products are promoted at the right time to the right audiences. Additionally, AI-driven strategies offer substantial benefits when it comes to efficiency and cost-effectiveness[6]. To plan, execute, and monitor traditional advertising campaigns, significant time and resources are often required. Artificial intelligence-powered tools, on the other hand, can automate many of these processes, reducing the need for manual intervention and allowing marketers to focus on strategic decisions. An example of this is the use of artificial intelligence algorithms in programmatic advertising platforms, which allow advertisers to automate the purchase and placement of ads in real-time based on consumer behavior and engagement metrics[7]. Automation not only streamlines the advertising process but also maximizes return on investment by allocating marketing budgets to the most effective channels and audiences[8]. AI has also been integrated into food marketing to enable more dynamic and responsive consumer engagement. In traditional advertising techniques, schedules and content are fixed, which limits their ability to adapt to consumer preferences and market conditions as they change[9]. AI-driven strategies, on the other hand, can continuously learn from consumer interactions and adjust marketing messages as needed. Marketers can take advantage of this adaptability to respond to emerging trends, hear from consumers, and refine their campaigns in order to achieve better results. A chatbot powered by artificial intelligence can engage with consumers on social media platforms, offering personalized recommendations, answering queries, and collecting valuable insights that can be used to improve future marketing campaigns[10].

Marketers can access large volumes of consumer data using AI technologies to gain deeper insights into their behavior. It can be difficult to identify the nuances of individual consumer preferences when traditional advertising techniques rely on aggregated data and broad market research[11]. AI-driven analytics, on the other hand, can analyze granular data from multiple sources and reveal patterns and correlations that would otherwise be difficult to spot. Marketers can use these insights to create more accurate customer profiles, segment their audiences more effectively, and design highly targeted campaigns using these insights. AI-driven strategies present numerous advantages, but they also present certain challenges and ethical considerations. The privacy and security of data are among the primary concerns[3]. Food marketing relies heavily on access to large amounts of consumer data, raising questions about how this information is collected, stored, and used. Developing consumer trust requires marketers to comply with data protection regulations and adopt transparent practices[12]. The sheer volume of targeted marketing messages could also lead to over personalization with AI-driven strategies. In order to maintain consumer trust and engagement, a balance must be struck between personalization and privacy. The implementation of AI in food marketing also requires significant investments in technology and expertise[13]. Cost savings and improved efficiency can be achieved with AI-driven strategies, but the initial setup costs can be high. For marketers to stay on top of AI advancements, they must invest in advanced analytics tools, hire skilled data scientists, and update their technology frequently. A necessity for ongoing training and education is also necessitated by the complexity of AI algorithms. AI is transforming how brands connect with consumers through food marketing. Compared to traditional advertising methods, AI-driven strategies provide an unprecedented ability to personalize recommendations, predict consumer behavior, and optimize marketing efforts[14]. The successful implementation of AI requires careful consideration of data privacy, ethical considerations, and the necessary investments in technology and expertise. As AI continues to evolve, it will undoubtedly play a more important role in shaping the future of food marketing, allowing brands to create marketing campaigns that are relevant, engaging, and effective. Traditional advertising techniques and AI-driven strategies are compared in detail in the study, highlighting their strengths and limitations[15]. This study reviews how artificial intelligence technologies, such as machine learning and natural language processing, are revolutionizing the food marketing industry. By examining how personalized recommendation systems and predictive analytics can be applied in food marketing, the study sheds light on how these AI-driven tools can enhance consumer engagement, satisfaction, and conversion. AI is able to create personalized marketing campaigns that resonate with individual consumer preferences. Comparing AI-driven marketing strategies to traditional marketing methods, the study evaluates their efficiency and cost-effectiveness. Using AI-powered automation and real-time data analysis, it streamlines marketing processes, reduces manual intervention, and optimizes budget allocation, resulting in higher returns. Marketers seeking to incorporate AI into their food marketing strategies should take action based on the study's findings. Investing in AI technologies, training personnel, and continuously updating marketing approaches helps companies stay competitive in an increasingly digital landscape by providing guidance. The study identifies areas for future research, such as the long-term impacts of AI-driven marketing on consumer behavior and the potential for over personalization. AI can be leveraged in food marketing to address emerging challenges and opportunities.

## 2. Literature review

Food marketing has been significantly transformed by artificial intelligence (AI). A new generation of AI-driven marketing strategies offers unprecedented levels of personalization and predictive accuracy, replacing traditional advertising techniques. In this literature review, we examine how AI can enhance personalized recommendations and predictive analytics in food marketing. In this review, traditional advertising techniques are compared with AI-driven strategies in order to provide a comprehensive understanding of the potential benefits and challenges of integrating AI into food marketing. Key studies in this field examine the use of generative AI for creating realistic marketing texts[16], the impact of AI on consumer behavior in the pet food market[5], and the implications of AI-generated food images on consumer perception[6]. This comparative analysis highlights AI's potential for improving marketing effectiveness, consumer engagement, and market dynamics. Camelo et al. [17]examined how sustainable development is integrated within Portugal's agri-food sector, emphasizing the importance of promoting sustainable practices among businesses by integrating Sustainable Development Goals (SDGs) and environmental, social, and governance (ESG) criteria. Based on qualitative analysis and comprehensive interviews with managers and experts, this study revealed a disparity between SMEs and larger companies in terms of the maturity of sustainability practices. As a result of a limited budget and external pressures, SMEs often take a reactive approach to sustainability, while larger companies tend to take a proactive approach to sustainability. Sustainability practices must be embedded into organizational culture and strategy, according to the study.

Hamdy et al. [18] reported that tourists' experiences influence the indirect relationships between extrinsic motivations, intrinsic motivations, and perceived destination images. This study used structural equation modeling with data from 613 international tourists of various nationalities to test five hypotheses. The findings indicated that second-order destinations' extrinsic motivations directly impacted tourists' intrinsic motivations and perceived destination image. Additionally, tourists' experiences moderated the direct effect of DEM on PDI for first-time visitors compared to repeat visitors and enhanced the direct effect of TIM on PDI. Kuang et al. [16] examined the challenges and practicality of using Natural Language Generation (NLG) to generate marketing messages in the Food and Beverage (F&B) industry. The study evaluated the effectiveness of existing techniques, such as Long Short-Term Memory (LSTM) and Open Pretrained Transformer (OPT), for creating realistic marketing texts. Califano et al. [6]found that consumers could distinguish between AI-generated images of food and authentic images when the information was disclosed. Participants found AI-generated images easy to recognize, particularly those of ultra-processed foods. When AI-generated images were not disclosed, they were less appealing than authentic ones, but when identified as artificial, their appeal nearly equaled that of authentic pictures.

Dalgic et al. [19]conducted a conceptual study to identify technologies that could mitigate the shortage of human resources at events and determine areas where these technologies might be useful. The study concluded that robots/robotics, artificial intelligence, the Internet of Things, and augmented reality/virtual reality could replace human resources in events due to shortages. Jayanthi & Shanthi [20]reviewed the use of machine learning and deep learning algorithms for predicting crop yields. The systematic literature review identified common methods and features used in crop yield prediction studies. The review found that climate variables such as temperature, humidity, wind velocity, rainfall, sunlight hours, water level, and soil moisture were frequently used in crop yield predictions. Furthermore, deep learning algorithms outperformed traditional machine learning algorithms. Upendra et al. [21] investigated the impact of the COVID-19 pandemic on agriculture and food production in India, noting that the lockdown significantly affected these sectors. During the initial phase of the lockdown, Rabi crops were in the harvesting stage, leading to supply chain disruptions and reduced marketability. The study underscores the importance of understanding the impacts of global pandemics on agriculture and food production. Liu et al. [5]examined consumer behavior in the pet food market, focusing on factors influencing online and offline purchases. A comprehensive survey of pet owners and potential consumers analyzed the impact of price, logistics, service, and quality on decision-making. The findings indicated that price sensitivity significantly influenced consumer choices, with competitive pricing driving online purchases as consumers sought cost-effective options. Dulhare et al. [22] examined the use of AI and UAVs in rice cultivation, from plowing to harvesting. The study aimed to predict soil fertility and detect diseases, pests, and weeds early to increase yields. Manual detection methods are labor-intensive and time-consuming. To address this, AI algorithms were used to classify images of soil fertility using UAVs. The focus was on automating rice cultivation to reduce pesticide use and improve efficiency. Li et al.[23] introduced D-AdFeed, a location-based advertising framework that considers diversity when assigning ads to users. Existing systems often send users ads with the highest utility scores without considering category diversity, leading to repetitive and uninteresting advertisements. D-AdFeed addresses this by balancing ad utility with user satisfaction using a diversity-aware component.

Ssematimba et al. [24] examined the geographic location and pathways of influenza A virus infection on commercial upland game bird farms in the United States. The study aimed to provide evidence supporting new strategies to minimize avian influenza outbreaks. The findings highlighted specific pathways for upland game bird farms, differing from those targeting poultry farms producing chickens, turkeys, and eggs, emphasizing the need for tailored control measures to effectively minimize AI outbreaks. Wetherill et al. [10] investigated food choice considerations among American Indians in rural Oklahoma. They administered the Food Choice Values Questionnaire to 83 participants who frequently shopped at tribally owned convenience stores. The results showed that sensory appeal, cost, and health were significant factors influencing food choices. Roberts et al. [13] examined the reproductive performance of heifers provided with either unrestricted or limited access to feed after weaning. The study found that heifers fed at 80% of their appetite-adjusted body weight had lower average daily gains and consumed less feed compared to those with ad libitum access. The reproductive performance was evaluated in composite heifers born over three years, randomly assigned to either the control group (fed ad libitum; n = 205) or restricted feeding for 140 days. Aguado et al. assessed the effectiveness of the LAMDA classifier in identifying customer behavior and predicting which customers are most likely to defect when a new retailer enters the market. As shown in Table 2, we present a summary of recent studies on the use of AI and technological developments across diverse sectors.

**Table 1:** Summary of Recent Studies on AI and Technological Applications Across Various Sectors

| Authors | Year | Method | Result |
|---|---|---|---|
| Camelo et al.[17] | 2024 | Interviews and qualitative analysis | A disparity exists between the maturity of sustainability practices in SMEs and larger companies, with SMEs taking a reactive approach due to low budgets and external pressures, whereas larger companies adopt a proactive approach. |
| Hamdy et al.[18] | 2024 | Modeling structural equations | Observations of tourists effect indirect relationships between extrinsic and intrinsic motivations, and the perception of a destination, with moderate effects for a first-time visitor and a repeat visitor. |
| Kuang et al.[16] | 2024 | Techniques for evaluating NLG | The F&B industry can benefit from existing techniques like LSTM and OPT for creating realistic marketing texts. |
| Califano et al.[6] | 2024 | Comparative analysis | Images generated by AI can be distinguished from authentic food images; disclosure makes AI-generated images more appealing. |
| Dalgic et al.[19] | 2024 | Study of concepts | Events can be mitigated as a result of the use of robots, artificial intelligence, IoT, and augmented reality/virtual reality. |
| Jayanthi & Shanthi[20] | 2024 | Literature review in a systematic manner | Using climate variables, deep learning algorithms outperform traditional machine learning algorithms. |

| Richards et al.[25] | 2024 | Examination of curation modes | Destination marketing organizations are increasingly curating tourism products, providing branded hospitality areas, and curating places. |
| --- | --- | --- | --- |
| Dutta et al.[26] | 2023 | Transfer learning with zero shots | It is possible to predict optimal ripeness levels for a variety of climacteric fruits using generic AI models even when there are no labeled data to be found. |
| Dhillon et al.[27] | 2023 | Barriers and opportunities review | Highlighted potential opportunities provided by technological advancements for small-scale farmers. |
| Monaco et al.[28] | 2023 | Metaverse analysis | Analyzed potential benefits and challenges of the Metaverse for tourism and research sectors. |
| Chiras et al.[7] | 2023 | Models for explaining machine learning | In Northern Greece, food-related lifestyle plays a key role in shaping consumer behavior regarding chicken meat consumption. |
| Xie et al.[29] | 2023 | Review of robotic grasping techniques | It is imperative that logistics, FMCG, and food delivery sectors have adequate learning policies for flexible object grasping. |
| Upendra et al.[21] | 2023 | Impact analysis | The COVID-19 lockdown significantly affected agriculture and food production in India, especially during the Rabi harvest. |
| Woo et al.[30] | 2022 | Systematic search | Predicting current development trends in Software as a Medical Device (SaMD) and Digital Therapeutics (DTx). |
| Philp et al.[4] | 2022 | Google Vision AI analysis | Instagram users engage more with food images that appear typical, suggesting that easier mental processing drives positive feelings. |
| Dulhare et al.[22] | 2022 | Image classification algorithms using artificial intelligence | Using artificial intelligence and unmanned aerial vehicles to automate rice cultivation, predict soil fertility, and detect diseases, pests, and weeds early. |
| Singh et al.[31] | 2021 | COVID-19 lockdown analysis | The delivery of livestock services in Uttar Pradesh, India is declining except for milk consumption. |
| Wu & Chien[32] | 2021 | AI deep-learning method with two stages | Assessed the similarity between online product illustrations and the actual goods received offline to mitigate quality risks in omnichannel operations. |
| Chiu & Chuang[33] | 2021 | Transfer learning application | On a sharing kitchen platform, focused on AI refinement in chatbots for precision marketing. |

| Coope et al.[34] | 2021 | Hydrocortisone granules | As a result of Alkindi's work, pediatric adrenal insufficiency is now being treated with an age-adjusted hydrocortisone preparation. |
| Li et al.[35] | 2021 | Framework for location-based advertising | Utilizing a diversity-aware component, D-AdFeed balances ad utility and user satisfaction. |
| Ssematimba et al.[24] | 2019 | Analysis of geographic locations and pathways | Researchers have identified specific pathways for the infection of upland game birds with the avian influenza virus, highlighting the need for tailored control measures. |
| Rahman et al.[36] | 2019 | Observations and direct interviews | Researched two management systems for semi-arid dairy buffalo production (HYI and completely intensive). |
| Schirra et al.[37] | 2005 | Comparative analysis | IMZ and FLU effectively controlled citrus fruit decay; residue levels varied with FLU concentration and temperature, affecting epicuticular wax ultrastructure. |
| Yee et al.[38] | 2008 | LPAI control program analysis | LPAI outbreaks in LBMs were prevented through active surveillance, prevention, and rapid response. |

## 3. Market Orientation and Its Measurement

### *3.1 The Concept of Market Orientation*

Marketers have long asserted that market orientation is a crucial driver of business success. Since the 1950s, the concept has been utilized in various forms within marketing and management literature. However, significant progress in defining market orientation and providing empirical evidence of its impact on business performance has only been achieved recently. Market orientation encompasses gathering market intelligence about current and future customer needs, disseminating this intelligence across departments, and responding accordingly. Empirical research indicates that better business performance is correlated with higher levels of market orientation. Notably, market turbulence, technological changes, and competitive intensity do not significantly affect the strength of this relationship.

## 4. Interactive Digital Marketing

Digital interactive marketing (IDM) involves a variety of communication tools and platforms, including text messaging and visual media, social networking sites, product review websites, wikis, blogs, chat rooms, online gaming sites, and websites that host user-generated content like videos, photos, and consumer reviews. Marketing-relevant personal data is collected and processed by digital technologies. Consumer engagement can be improved by aggregating this data and processing it at a micro level, such as customized purchase incentives and online behavioral advertising. Conversational and collaborative marketing is replacing traditional, centralized, and distributed marketing methods.

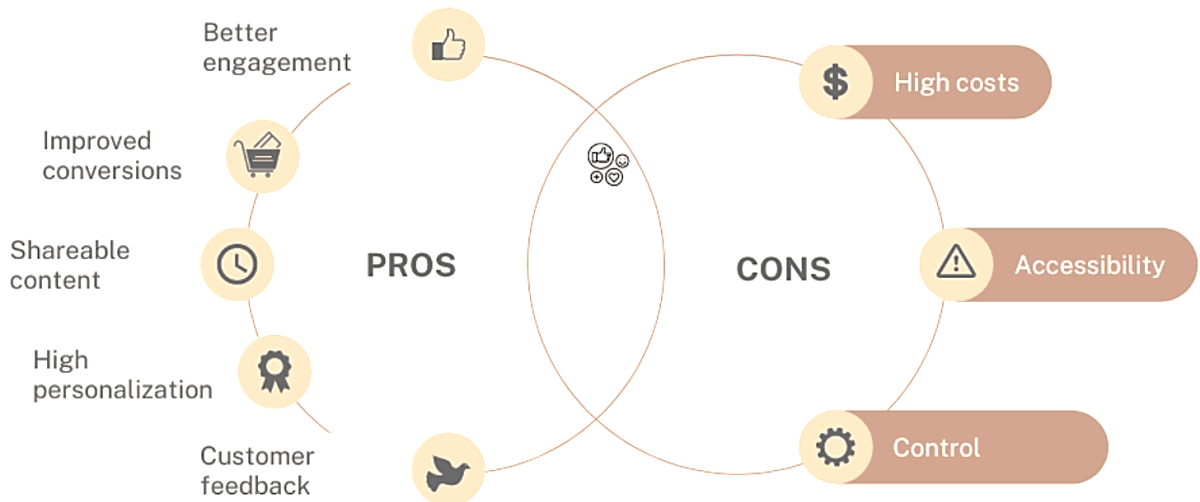

**Figure 1:** Advantages and Disadvantages of Interactive Marketing

**Table 2:** Summary of Recent Studies on AI and Technological Applications Across Various Sectors Including Digital Marketing in the Food Industry

| Authors | Year | Method | Result |
|---|---|---|---|
| Califano et al.[6] | 2024 | Experiments on the internet | The participants were able to easily recognize AI-generated food images, especially ultraprocessed foods. These images became more appealing after the AI generation was revealed. |

| Ding et al.[39] | 2023 | AI and big data analysis | In the food industry, artificial intelligence and big data enhance food safety, production, and marketing by driving innovation and meeting consumer needs. |
| --- | --- | --- | --- |
| Liu et al.[5] | 2023 | Comprehensive survey | Competitive pricing drives online purchases in the pet food market, influencing consumer behavior significantly. |
| Monaco et al.[28] | 2023 | Metaverse analysis | Researchers identified potential benefits and challenges of the Metaverse in tourism, emphasizing the need for further research into the social, economic, and environmental impact of the Metaverse. |
| Nordmark et al.[40] | 2021 | System development for decision support | As a result of IoT and AI technologies, Swedish apple and grapevine production is more sustainable and competitive. |
| Rowan et al.[1] | 2023 | Analyzing digital technologies | Sustainable fishery and aquaculture requires digital transformation, which improves production, selection of species, and disease mitigation, while reducing waste and pollution. |
| Sharma et al.[41] | 2021 | AI and big data in the food industry | Food industry innovations driven by artificial intelligence and big data analytics improve logistics, supply chain management, marketing, and production. |
| Woo et al.[30] | 2022 | Analysis of regulatory science programs | In addition to funding for SaMD and DTx development, MFDS' Innovative Medical Device Program also supports the development of these devices, providing more options for patients. |
| Yaiprasert et al.[3] | 2023 | Analyses of ensemble machine learning | The naive Bayes algorithm achieved 97.18 percent accuracy in digital marketing for food delivery businesses using decision trees and nearest neighbors. |

In Table 2, we present a summary of recent studies on AI and technological applications in various sectors, such as digital marketing in food. Figure 1 is a Venn diagram that illustrates the advantages and disadvantages of interactive marketing. On the left, the pros include better engagement, improved conversions, shareable content, high personalization, and customer feedback. On the right, the cons highlight high costs, accessibility issues, and control challenges.

There is a mix of brand-generated content and spontaneous commentary in these conversations. Commercial sectors face substantial risks as well as benefits from such interactions. Almost seventy-five percent of respondents to the Nielsen market research company's 2007 global survey believed peer-to-peer recommendations were the most trusted form of advertising. As a result of the digital dissemination of information and opinions, brands are subject to public scrutiny and commentary. Reputational damage can be caused quickly by negative reactions. By minimizing these risks, companies avoid the perception of manipulation. User-generated content and secondary social responses have already been monitored and evaluated, independently of confidential commercial strategies.

**Table 3:** An introduction to artificial intelligence concepts

| Term | Definition |
|---|---|
| Artificial Intelligence | Making a machine behave in ways that would be considered intelligent if a human were so behaving. |
| Machine Learning | A computer program learns from experience E with respect to some tasks T and performance measure P if its performance at tasks in T, as measured by P, improves with experience E |
| Deep Learning | Methods for representation-learning that have multiple levels of representation, resulting from the composition of simple but non-linear modules that transform one level of representation into a higher level of abstraction |

Table 3 outlines the definitions of essential concepts in the realm of artificial intelligence, including Artificial Intelligence (AI), Machine Learning (ML), and Deep Learning (DL). Artificial intelligence (AI) has been around since antiquity, with research beginning in the 1950s. During the Dartmouth Summer Research Project on Artificial Intelligence held in 1956, John McCarthy coined the term Artificial Intelligence and asked, "Can machines think?" (1950). Alan Turing's famous Turing Test posed the question, "Can machines think?" (Turing, 1950). AI refers to making a machine behave intelligently like a human. After focusing on math and logical reasoning problems in the 1980s, AI research shifted to knowledge and expert systems. In spite of their initial promise, both waves were challenged and failed to deliver the results expected.

## 5- Traditional advertising

Mass marketing communication, like radio and television, targets consumers by utilizing demographic information. In the past decade, the introduction of the Internet has led to a change in advertising, and digital advertising has emerged as a distinct form of advertising compared with traditional methods. It is a two-way communication model in digital advertising that allows consumers to communicate with brands through e-mail, text messages, and social media, increasing consumer loyalty through interaction. In published research, online and traditional media advertising have not been compared and therefore online advertising and other digital channels are not included as digital advertising. According to Zahradka et al.[42], a fully remote digital clinical study using Current Health (CH) was designed and conducted to collect symptoms and continuous data from COVID-19 participants. During the pandemic, a comprehensive digital protocol was deployed and assessed in a real-world scenario. As a result, the digital protocol proved feasible, efficient, and effective for collecting high-quality data.

A study conducted by Willis et al. [43]examined how social media platforms such as Twitter and Instagram influence health literacy about prescription medications. Patients who had previously collaborated with pharmaceutical companies were interviewed by the researchers. This study found that influencers shared personal experiences, explained medication side effects, and stressed the importance of adhering to treatment regimens as ways to communicate health literacy information about prescription medications. Sestino et al. [15] investigated consumers' reactions to lab-grown meat (LGM), a new green food created using tissue engineering. LGM has a comparable nutritional value to traditional meat, as well as potential sustainability benefits, so the study assessed the impacts of advertising and communication strategies. Findings highlight the importance of influencing consumer attitudes toward LGM. As a result of these results, marketers can better promote sustainable consumption behaviors through innovative food products like LGM. According to Gaspar et al.[44], traditional agri-food companies in central Portugal can utilize Information and Communication Technologies (ICTs) to modernize and innovate their methods. According to a survey of 50 companies, most of them used computers and internet services for business purposes, including cereals, cheese, olive oil, dry sausages, honey, wine, and horticulture. In contrast, many businesses lacked a website and did not use the internet for advertising or sales.

According to Axenov et al.[45], e-grocery and ready-made food trade are spatially organized in a major Russian city. In this study, we sought to identify novel characteristics of food retail businesses that set them apart from traditional food retailers. The requirements imposed by online food retail on urban spaces were described in St. Petersburg by researchers. In order to demonstrate how e-grocery trade fits within the city's existing infrastructure, they analyzed the spatial and temporal aspects of this new shopping model and compared it with established models. Mazzolani et al. [46]investigated the relationship between disordered eating attitudes and food choice motives among vegans in Brazil using a cross-sectional online survey. In the study, vegan diets were linked to higher levels of disordered eating attitudes, especially those regarding body weight and shape. Healthy food choice motives were positively associated with intrinsic motivations (e.g., values), while extrinsic motivations (e.g., social pressure) were negatively associated. The Summary of Studies Investigating Traditional Advertising and Consumer Behavior can be found in Table 4.

**Table 4**: Summary of Studies Investigating Traditional Advertising and Consumer Behavior

| Author | Year | Method | Finding |
|---|---|---|---|
| Amerzadeh et al.[47] | 2022 | Qualitative study with interviews of 30 key stakeholders | A lack of food education, cultural norms promoting high-calorie foods, and a lack of labeling regulations are identified as key factors in unhealthy consumption |
| Russo et al.[48] | 2022 | Neuromarketing approach | Consumers perceive products with certifications of origin as being more sustainable, safer, and of higher quality; proposed neuromarketing techniques for effective commercials |
| Dikmen et al.[49] | 2022 | Cross-sectional study using WHO's nutrient profiling model | Adolescents' food preferences are significantly influenced by television advertisements |
| Putra et al.[50] | 2023 | Analysis of regional tourism authorities' Instagram posts | Incorporated traditional Southeast Asian food images into tourism advertising to reflect cultural identity and highlight unique flavors |
| Ahmad et al.[51] | 2023 | Comparative study of perspectives in Islamic, European, and Malaysian contexts | Muslim consumers are highly aware of halal meat and food fraud has serious implications, as evidenced by the 2020 scandal involving false advertising |
| Czaplicki et al.[52] | 2022 | Examination of nicotine warning statements in ENDS advertisements | Pre- and post-regulation nicotine warning statements were investigated |
| Jibril et al.[53] | 2022 | Quantitative research design with 405 valid responses from fast-food customers | Among fast-food patrons' selections, convenience, time, cost, and taste and preference are the most important factors, moderated by traditional media advertisements and electronic word-of-mouth. |

As consumers make purchases in various channels, their preference for the channels they will use always changes throughout their lifetime. Marketers can use consumer-advertising preferences to target consumers and influence purchase behavior effectively when they understand consumer-advertising preferences.

## 6- Leveraging artificial intelligence to monitor unhealthy food and brand marketing to children on digital media

Commercial promotions for particular foods and brands are known as food and brand marketing. In several systematic reviews it has been shown that unhealthy food and brand marketing, especially in video games and on television, adversely affects children's health and diet quality. Marketing has traditionally involved one-way communication of information during certain times and places (e.g., television commercials). However, digital media—including overt and covert (eg, product placement, social media influencers) promotional activities on websites, social media, text messaging, applications, email, and online games—allows marketers to push unprecedented volumes of information to children in real time, often using artificial intelligence (AI)-enabled tactics.4,5 In the digital age, commercial messages no longer interrupt, but are instead intimately integrated with content,6 and thus cues that can help children identify marketing (eg, commercial breaks) are absent or not prominently displayed.

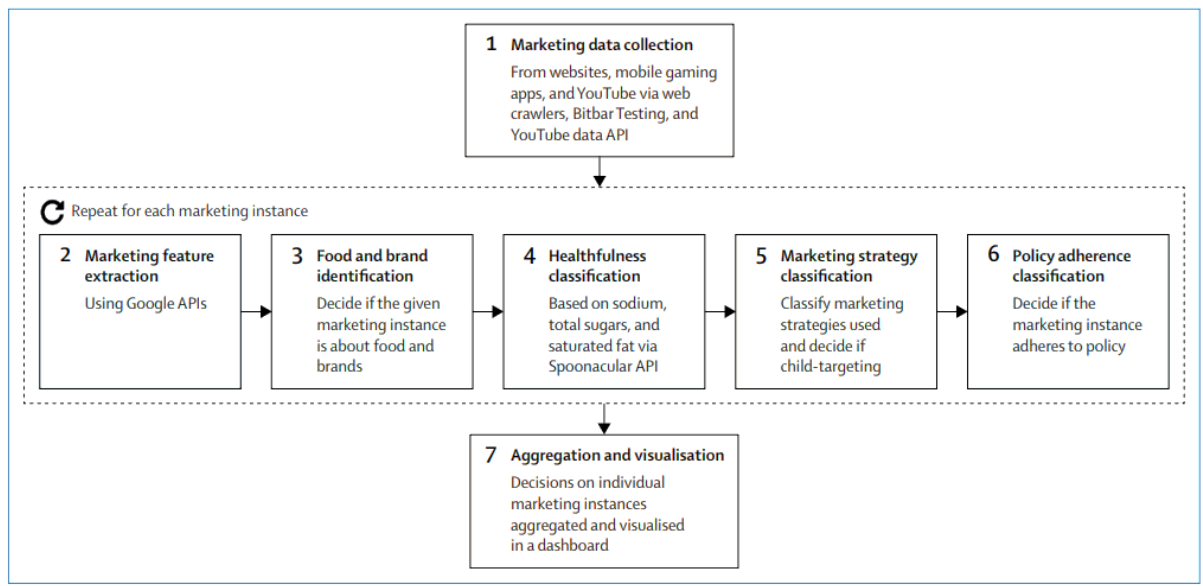

**Figure 2:** depicts the workflow for the AI-driven analysis and classification of marketing strategies.

An AI-driven marketing analysis follows a step-by-step process beginning with data collection (websites, mobile apps, YouTube) and continuing with marketing feature extraction using Google APIs. This figure 2 outlines the step-by-step process for identifying food and brands, identifying healthfulness (using nutritional data from Spiracular API), categorizing marketing strategies, categorizing policy adherence, and finally aggregating and visualizing the results. Throughout the process, each step feeds into the next, ensuring a comprehensive analysis and formulation of a strategy. This makes it more difficult for children to recognize marketing on websites than on television. Because of this, digital marketing may foster more sustained and deeper engagement with unhealthy foods and brands, with potentially more negative impacts on health and diet.

## 7. Motivation and Background of the Research

### *7.1. Implications of AI Models for Decision Making in Agriculture*

Artificial intelligence (AI) models are becoming increasingly important for agricultural decision-making [23]. With this shift, digital marketing moves from an operational role to a primary strategic component within a decision-making framework [24,25]. The use of AI models, in particular artificial neural networks (ANNs), is essential for precision agriculture because it allows large datasets to be processed effectively [26]. Efficiencies in resource allocation and marketing costs are improved through this process [27,28]. Dynamic resource optimization is facilitated through AI models, especially artificial neural networks [29,30], emphasizing the intricate relationship between operational decisions and marketing success. ANNs, in particular, play a significant role in early disease detection, which is crucial to the success of digital marketing initiatives [31]. As a result, not only are crop yields protected, but also economic indices such as the return on investment for digital marketing and the cost-effectiveness of disease management strategies are affected [32,33,34]. AI models, including ANNs, enhance agricultural enterprises' competitiveness in terms of market trends and price forecasting [35,36]. Economic indices linked to market trends and pricing dynamics also play a role in strategic decisions related to digital marketing campaigns. Agriculture-specific decision support systems (DSS) have been transformed by the integration of AI models, specifically ANNs. Agricultural decision-making is strongly affected by economic indicators related to digital marketing strategies [24].

**Table 5:** Summary of Studies Investigating AI, Big Data, and Marketing in Agriculture

| Author | Year | Method | Finding |
|---|---|---|---|
| Ding et al.[39] | 2023 | Food industry AI and big data analytics review | AI and big data have become essential for enhancing food safety, production, and marketing, with the potential for further advancements as technology evolves. |
| Giannakopoulos et al.[54] | 2024 | Analyses of digital marketing analytics and agroeconomic indicators | Digital marketing analytics, combined with AI-based modeling, can improve decision-making processes in agricultural firms, helping to alleviate financial difficulties. |
| Hackfort et al.[55] | 2024 | A study of agricultural corporate strategies | Leading input supply and machinery companies in agriculture create economic and non-economic value from big data, contributing to market concentration and monopolization, with significant socio-ecological implications. |
| Singh et al.[56] | 2019 | A case study of Agroy, an Indian manufacturer of agri-products | The study highlighted the challenges in marketing within rural India, including low technological awareness, limited digital transactions, and language barriers. [Insert specific findings]. |
| Sarker et al.[57] | 2020 | Digital farming and big data | Big data technologies can assist farmers in making real-time informed decisions, though adoption is limited due to infrastructure needs, expertise, technological knowledge, and awareness. |
| Jayashankar et al.[58] | 2020 | Big data and co-creation in agriculture | Farmers gain monetary and non-monetary value from interactions with agricultural companies, autonomous decision-making, and episodic events through the use of big data technology. |
| Mao et al.[59] | 2024 | Cross-border e-commerce in China's green food industry | Cross-border e-commerce growth demands higher standards for data transmission and exchange, impacting the green food industry chain. |
| Yigitcanlar et al.[60] | 2020 | A study of AI in urban planning and development in Australia | AI is perceived as a versatile technology with applications in marketing, banking, finance, agriculture, healthcare, security, space exploration, robotics, transport, chatbots, and manufacturing. |

AI models, particularly ANNs, enhance digital marketing activities by uncovering complex relationships between agroeconomic indices and key website metrics. In order to refine and optimize digital marketing strategies for agriculture firms, it is important to have a nuanced understanding of how digital marketing affects sectoral decision-making and economic indexes. A final note on AI models, especially ANNs, is that they influence agriculture decision-making beyond the advances brought by technology [41]. For informed and impactful marketing decisions, strategic integration is necessary.

### 7.2 Big Data and Digital Marketing in Agriculture

Marketing strategies are heavily influenced by agroeconomic indicators, which are essential tools for navigating a constantly changing economic environment [42]. Indicators of the broader economic environment and consumer behavior can provide marketers with useful insights [43]. During economic expansions and contractions, these indicators impact advertising budgets and resource allocation, allowing for adjustments based on inflation rates and economic stability. Through adjustments to digital marketing strategies, agroeconomic indicators can be observed and predicted. Agroeconomic indicators provide marketers with insights into consumer behavior, helping them to ensure messages are delivered effectively to a specific demographic [47]. As pricing strategies are adjusted in response to economic indicators, marketing campaigns are more agile, aligning with prevailing economic conditions [48]. Agricultural companies can influence their financial performance and decision-making processes through digital marketing strategies. Agroeconomic index variations could potentially be analyzed using these strategies. In addition to leveraging economic fluctuations strategically to optimize outcomes, marketers benefit from insights from economic indices in crafting strategies that withstand and withstand economic fluctuations. The influence of agroeconomic conditions extends seamlessly into digital marketing, significantly affecting consumer behavior and shaping strategic initiatives online [49]. Particularly during economic downturns, marketers adopt a strategic approach to website optimization, emphasizing elements such as pricing transparency, tangible discounts, and value propositions designed to resonate with the cost-conscious sentiments of consumers [50]. This careful optimization aims not only to enhance the user experience but also to align the online presence of brands with the economic realities faced by consumers.

Users' experience (UX) is carefully tailored in website design and functionality in order to align with evolving expectations influenced by economic changes [51]. For marketers to meet the dynamic needs of consumers, navigation is optimized, loading times are reduced, and mobile responsiveness is ensured [52]. A thoughtful optimization of the user experience and a strategic response to changing economic conditions influence online consumer behavior through this method enhances the overall user experience. Beyond technical aspects of website design, agroeconomic factors influence various aspects of digital marketing. The digital marketing landscape is also significantly impacted by content marketing and search engine optimization (SEO). As economic conditions fluctuate, marketers need to adapt their SEO strategies to correspond with emerging search behaviors influenced by the current economic climate, deliver insights, and deliver insightful content [54]. As a result of this interdependence, marketers need to know agroeconomic indicators at a nuanced level [55], ensuring that strategies remain agile and responsive to the dynamic demands of users navigating the fluid and ever-changing economic landscape, particularly within the complex realm of digital marketing. As digital marketing activities in agriculture continue to evolve and users' online behaviors become increasingly complex, the integration of big data analytics becomes increasingly significant [12]. Stakeholders can gain sophisticated insights into a variety of dimensions using big data analytics, such as consumer behavior, market trends, and environmental impacts of agricultural practices [56]. Marketers need to use this data-driven approach to shape their marketing strategies and facilitate informed decisions in the agricultural supply chain. Agricultural sustainability initiatives are driven by the synergy between agroeconomic indicators, big data analytics, and digital marketing strategies. A combined approach not only enhances the precision of marketing strategies, it also aligns them with broader sustainability objectives. The integration of these key components fosters a holistic and data-driven approach to sustainable agriculture.

## 8. Collaborative Intelligence in Marketing

Various complementary methods of implementing collaborative AI are possible when AI possesses multiple intelligences. Human intelligence (HI) encompasses both marketer and consumer intelligence. For marketing and consumption tasks, AI of different intelligence levels can be utilized by both parties. Marketers and consumers are both included in this discussion of marketing. Both marketers and consumers can achieve collaborative intelligence in marketing by following three basic principles, along with sub-principles. The framework recognizes that AI is computational while HI is biological and identifies the relative strengths of each intelligence level. Following that, it describes how each level's strengths can complement the other for collaborative intelligence to be achieved. Finally, it reveals how collaboration can become more intelligent when it reaches a higher level of intelligence. A general principle of marketing is that AI and HI can each perform marketing tasks to the extent that each has a relative advantage based on their level of intelligence. The current AI is better at mechanical and analytical intelligence due to its computational nature, whereas the current HI is better at contextual, intuitive, and emotional intelligence due to its biological nature. The second general principle (GP2) builds on the first, suggesting that lower-level AI should enhance higher-level intelligence. There are three sub-principles (GP2a–2c) that explain how marketers and consumers can benefit from different AI intelligences. GP2a is an example of a mechanical AI augmenting contextual HI for undesirable marketing tasks (GP2b is an example of a mechanical and analytical AI augmenting intuitive HI for smarter marketing decisions (GP2c is an example of all AI intelligences supporting emotional HI for better emotional intelligence in marketing). In general principle 3 (GP3), artificial intelligence augments and then replaces human intelligence within each intelligence level. Following this principle, AI will eventually achieve autonomous performance of all lower-level intelligence tasks, pushing collaboration to a higher level of intelligence based on historical observations and future projections of AI's progression from mechanical to analytical to emotional intelligence.

### *8.1. AI and Computer Vision in the Food Industry*

Since recent innovations, artificial intelligence has experienced rapid growth, with experts predicting significant growth by 2020 [73]. Increasing digital data volumes will drive this growth, which is projected to reach 44 trillion gigabytes by 2020 [8]. Through innovative solutions to existing problems across a wide range of fields, huge amounts of data combined with trainable AI startups are fueling the 4th Industrial Revolution (4.0 IR). Machines and tools for building structures were introduced during the Industrial Revolution. 2nd Industrial Revolution (2.0 IR) began with the advent of electricity, radio, and airplanes. A new industrial revolution was ushered in by computers and the internet in the early 1970s. Several industries, including the food sector, are being impacted by recent advances in artificial intelligence (AI), computer vision, and big data [105]. Farming, cultivation, processing, and production have all been transformed by artificial intelligence in the food industry. It is now possible to access nutritional information and images of food on computers, in addition to displaying food images. As IBM's AI, Watson, suggested innovative recipes based on ingredients analyzed, it impressed renowned chefs with its ability to produce variations in recipes in 2016 [78]. Computer vision has been enabled by machine learning and deep learning. As of 2012, computers could understand and act upon image content through computer vision and image processing.

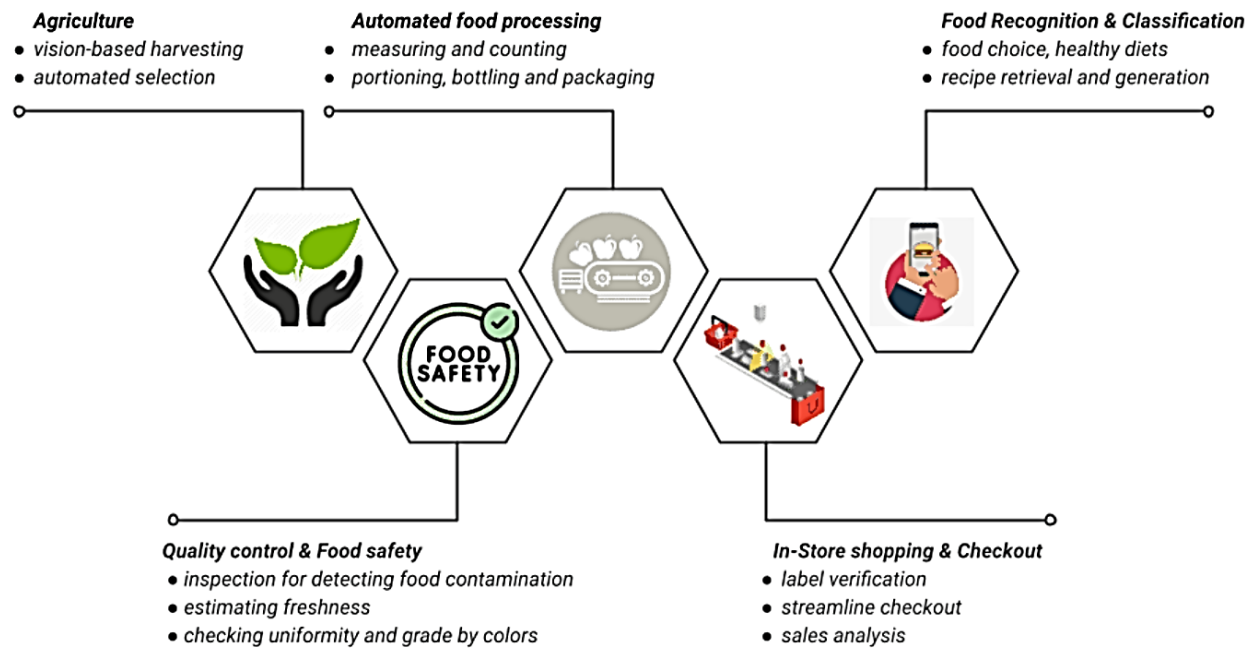

**Figure 3**: Applications of AI in Various Sectors of the Food Industry

This figure 3 illustrates the diverse applications of AI across different sectors, including agriculture (vision-based harvesting, automated selection), automated food processing (measuring and counting, portioning, bottling, and packaging), food recognition and classification (food choice, healthy diets, recipe retrieval and generation), quality control and food safety (inspection for detecting contamination, estimating freshness, checking uniformity and grade by colors), and in-store shopping and checkout (label verification, streamline checkout, sales analysis). In addition to object detection, recognition, tracking, and facial recognition, machine learning and deep learning have expanded computer vision's capabilities. As a result of deeper neural networks demonstrating superior performance in 2014, deep learning strategies are becoming increasingly important [97]. The performance of deep networks has been evaluated through benchmarks using large datasets for training and testing [37].

## 8.2. AI-Driven Customer Segmentation

A marketing strategy that is driven by AI will enhance accuracy and granularity by enhancing the granularity of customer segmentation. Large datasets, nuanced patterns, and subtle customer groups can be identified by AI algorithms. The deeper the insight, the more precise and effective the targeting. AI-driven segmentation is characterized by its dynamic nature. Artificial intelligence continuously learns and adapts to evolving customer behavior and market trends, keeping marketing strategies relevant over time by updating segments based on real-time market information. A key strength of AI in customer segmentation is predictive analytics; it uses past data to forecast future customer behavior and preferences, allowing companies to anticipate customer needs and tailor marketing efforts accordingly. Customer engagement and response rates are driven by personalization. As a result of AI, businesses can create highly targeted marketing messages that match the preferences of their target audiences at scale. Customer experiences are enhanced and relationships between businesses and clients are strengthened through this level of customization.

Segmentation driven by AI offers cost efficiency as well, reducing the need for manual analysis and reducing time and resource requirements. AI also optimizes marketing spend by directing resources to segments that are most likely to generate positive results, so that return on investment can be maximized. In summary, AI-driven customer segmentation improves accuracy, adapts dynamically, predicts future trends, enables scalability, and lowers costs. Businesses seeking a competitive edge in today's data-driven marketplace can benefit from this tool. Segmentation driven by artificial intelligence can enhance targeted marketing strategies, enhancing customer engagement and improving business outcomes through accuracy, adaptability, and efficiency.

## 8.3. Interactive and Media-Rich Environments

As social media and mobile devices proliferate, interaction between firms and consumers has dramatically increased, with information encoded in rich media formats such as text, images, and videos. Rich media content provides valuable insights into consumer perceptions and preferences, as well as brand positioning for companies. A competitive advantage involves creating informative and engaging content that enhances awareness, perception, and acceptance. Rich media content is natural for humans, but it poses challenges for machines. Machine learning-based AI tools can generate insights and prescribe solutions for these interactive and media-rich environments by learning less.

*8.4. Personalization and Targeting*

There has been an increase in the personalization of marketing. Machine learning methods are advancing large-scale, context-dependent personalization and targeting markets rich with data and digital channels. Microsegments are being segmented more finely, rather than a handful of large segments. Segmentation increasingly incorporates preferences and behaviors, including Facebook likes, Google searches, and mobile phone apps.

**Table 6:** Stimuli Examples (Carrot)

| | Original photo | AI version |
|---|---|---|
| **Unprocessed** | 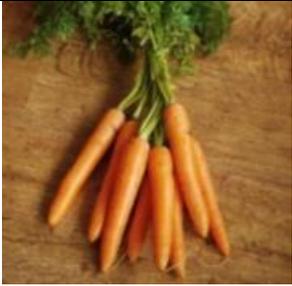 | 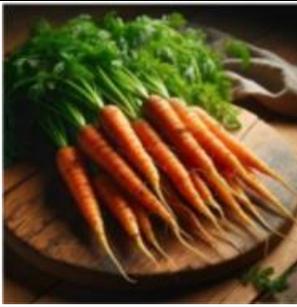 |
| **Processed** | 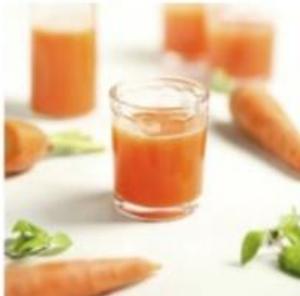 | 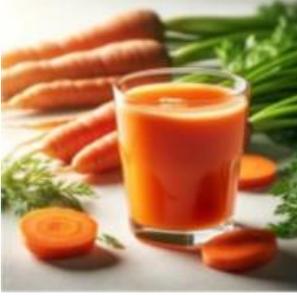 |
| **Ultra-processed** | 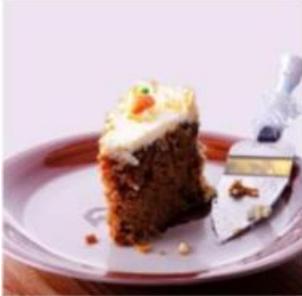 | 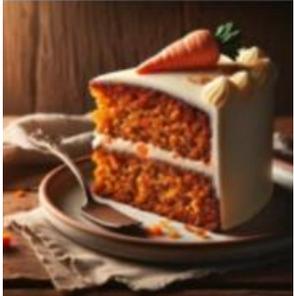 |

table 6 compares original photos and AI-generated versions of carrot images across three processing levels: unprocessed, processed, and ultra-processed. The unprocessed category shows fresh carrots, the processed category displays carrot juice, and the ultra-processed category features carrot cake. Personalized offers are tailored based on each consumer's individual profile because of continual refinement. There is a dynamic nature to preferences and needs, and there are temporary opportunities. Matching the right offering to the right consumer at the right time and in the right context is the key to effective targeting. Personalization and context-based targeting are largely driven by powerful machine learning algorithms, which are also driving the rapid evolution of these practices.

## 9. Comparing AI and Traditional Methods

### *9.1. Speed and Efficiency*

Ad creation powered by AI is significantly faster and more efficient than traditional methods. The process of creating a traditional advertisement requires considerable effort and time. From brainstorming to production and post-production, these processes can take weeks or even months. The use of artificial intelligence, on the other hand, allows for analysis of data, generation of content, and optimization of campaigns in real-time, which speeds up campaigns' launch. Data analysis using artificial intelligence allows for rapid insights and decisions. Machine learning algorithms can analyze audience data, predict performance outcomes, and make optimization suggestions almost instantly. Modifications can be made on the fly to ensure that a campaign is always performing at its best. In the fast-paced world of digital marketing, speed and agility are essential.

### *9.2 Cost-Effectiveness*

As far as cost is concerned, AI has a distinct advantage. Hiring creative talent and conducting market research are some of the costs involved in creating and distributing traditional advertisements. The high cost of these expenses often makes it difficult for small businesses to compete. AI-based tools, however, can automate many of these tasks, thereby reducing human resources. With automated content generation, real-time optimization, and predictive analytics, traditional methods can be reduced to a fraction of their cost. Thanks to this democratization of ad creation, businesses of all sizes can now create effective campaigns without breaking the bank.

### 9.3. Scalability and Reach

Scalability is another area where AI outperforms traditional methods. Creating personalized content for large audiences has always been challenging in traditional advertising. Customizing messages for different segments requires significant time and effort, resulting in one-size-fits-all campaigns that are often generic. Artificial intelligence allows for scalable personalization. Through the use of artificial intelligence, custom content can be generated for each user based on their individual data, ensuring each user receives engaging and relevant content. Scalability is not just about personalization. By combining AI with multiple channels and platforms, campaigns can be managed and optimized to maximize their reach and impact.

### 9.4. Accuracy in Targeting and Personalization

Targeting and personalization are vital to the success of any advertising campaign. Traditionally, market research and demographic data are used to identify target audiences. It is impossible to achieve true personalization with these techniques due to their lack of precision and granularity. The possibilities for targeting and personalizing are endless when you use artificial intelligence. Artificial intelligence can analyze a huge amount of data to determine whether a user will engage with an advertisement based on their behavior, preferences, and patterns. Precision targeting ensures that the right message reaches the right person at the right time. One aspect of personalization is demographics. The factors include browsing history, purchase behavior, and even real-time transaction context.

### 9.5. Creative Potential and Limitations

Analyze AI-powered ads' efficiency, cost-effectiveness, scalability, and accuracy by considering both their creative potential and limitations. Traditional methods are known for their ability to generate original, human-centric ideas that have emotional resonance. Artificial intelligence cannot replicate intuition, cultural understanding, and a deep connection with the audience that are often incorporated into the creative process.

# 10. Advantages of AI Marketing

## 10.1. Personalization

The ability to customize marketing experiences based on customer needs, interests, and preferences is one of the biggest advantages of AI marketing. Each customer's behavior, demographics, and purchase history are collected and analyzed. Artificial intelligence systems can use this data to automatically recommend tailored products, services, and content to customers based on their individual needs. The personalized approach significantly enhances customer loyalty and engagement, resulting in a more humane shopping experience.

## 10.2. Exact Targeting

Targeting is the process of precisely directing marketing efforts towards a specific audience in AI marketing. A consumer's interests, preferences, and buying habits can be identified using data analytics and behavior tracking. Marketers can personalize content and promotions for individual customers, increasing conversion rates. Furthermore, AI can be used to account for external factors, such as location and time of day, making it possible to refine targeting even further, resulting in more effective marketing and a more satisfying shopping experience for consumers.

## 10.3 Predictive Analysis

AI marketing predicts future events using data. Predictive analysis determines future outcomes by identifying patterns in data using machine learning algorithms. Instead of guesswork, marketers can make informed decisions using real-time data. Marketers can gain a competitive advantage by forecasting customer behavior, market trends, and product demand using predictive analysis. Businesses can increase sales and revenue by understanding customer preferences and tailoring their marketing campaigns accordingly. Predictive analysis, however, has limitations. Natural disasters and political upheavals may influence consumer behavior unexpectedly. A marketing campaign may also fail due to inaccurate data or flawed algorithms. It remains a powerful tool, however, when combined with other marketing strategies, which can help companies stay ahead of the curve and succeed in their marketing efforts.

# 11. Advantages of Traditional Marketing

## *11.1 Tangible Marketing Materials*

Marketing materials that are tangible are printed items that customers can touch and feel, such as brochures, flyers, and business cards. Traditional marketing methods often use these materials to leave a lasting impression on customers and provide them with something they can refer to later. In addition to showcasing a company's products or services, they can highlight its key features and highlight its advantages. Furthermore, tangible materials reinforce branding efforts and establish a cohesive visual identity across all marketing channels. In addition to face-to-face interactions, tangible marketing materials can be particularly effective at trade shows or networking events, since they provide potential customers with something tangible to take away from the encounter, which facilitates their decision-making. The use of these materials can help customers create a more memorable experience in a digitally dominated world. Digital marketing methods may not be as measurable as traditional marketing approaches, but they serve a distinct purpose.

## *11.2. Face-to-Face Interactions*

Communication between individuals occurs directly through verbal and nonverbal means during face-to-face interactions. In person marketing involves meeting potential customers, visiting their homes or businesses, setting up booths at fairs and events, or conducting one-on-one conversations. Through face-to-face interactions, marketers can convey brand messaging and build trust and relationships with customers. As a result, customers can ask questions and receive immediate answers, which builds trust and confidence. By picking up on verbal and nonverbal cues, marketers can also create emotional connections with customers. Face-to-face interactions can, however, be time-consuming and may not be as scalable as other marketing methods. A trained individual must be able to communicate effectively in person. Businesses looking to build strong relationships with their customers can leverage face-to-face interactions as a powerful marketing tool.

*11.3. Emotional Connections*

A brand's ability to establish an emotional connection with its customers is referred to as an emotional connection in traditional marketing. An in-person event, product demonstration, or sales pitch are often the means of forming this bond. In traditional marketing, relationships are built through personal engagement, emotional appeal, and engaging with customers on a personal level. AI marketing can't replicate this emotional connection because it lacks the human touch that traditional marketing provides. While AI marketing can offer personalized messaging and targeted advertising, it cannot convey the same level of empathy and understanding as human interaction. In order to build trust and credibility for a brand, emotional connections are essential. Customers often make purchasing decisions based on recommendations and testimonials from trusted sources. Traditional marketers can create loyal brand advocates by building emotional connections with customers.Traditional marketing relies heavily on emotional connections to establish brand loyalty, build trust, and create long-term customer relationships. Even though AI marketing is more efficient and personal, it still lacks the emotional touch of traditional marketing. A successful marketing strategy may require a combination of both methods.

## 12. Disadvantages of AI Marketing

*12.1 Lack of Emotional Connection*

One major disadvantage of AI marketing is the lack of emotional connection. While AI can analyze data and provide personalized suggestions, it cannot replicate the human touch that traditional marketing offers. For example, a virtual assistant may respond to customer inquiries in real-time, but it lacks the empathy, understanding, and emotional intelligence that a human customer service representative can provide. Although AI can incorporate some emotional elements through techniques like sentiment analysis, which helps identify how people feel about a brand or product, it still falls short in creating a deep emotional connection. To sum up, the lack of emotional connection in AI marketing is a significant challenge. While AI offers valuable analytical insights, it cannot replace the human touch of traditional marketing. Companies can leverage the benefits of both approaches by combining them effectively to mitigate this disadvantage.

*12.2 Dependency on Technology*

Dependency on technology refers to the reliance on artificial intelligence tools and software to complete marketing tasks. This reliance can lead to issues when there are errors or malfunctions, which can be damaging to the marketing strategy. It also means that marketers may lose the ability to conduct certain tasks manually, leading to a lack of flexibility and creativity. Additionally, heavy use of technology can reduce personal interactions with customers, potentially harming customer relationships. Overall, while AI offers many benefits, it is crucial for marketers to find a balance between using technology and maintaining human connections to avoid over-dependence.

*12.3 Decreased Creativity*

Another disadvantage of AI marketing is the potential for decreased creativity. AI marketing relies heavily on data analysis and predetermined algorithms to determine the best marketing strategies, which can limit a brand's ability to think outside the box and develop unique or unexpected ideas. While machine learning and AI can process vast amounts of data and provide insights, they cannot replicate human imagination. AI lacks the ability to think creatively or generate new ideas independently. It can only produce solutions based on existing data, which may result in marketing campaigns that do not stand out or are not as memorable as those created by humans. Despite this, AI can still produce predictable results, which can be useful in certain contexts. However, to maintain creativity and innovation, it is essential to integrate human input into AI-driven marketing strategies.

## 13. Disadvantages of Traditional Marketing

*13.1. Limited Targeting*

In traditional marketing, it can be difficult to focus a campaign precisely on a specific audience. As a result, resources can be wasted, and messages can be ineffective. There are several reasons for this limitation, including: Lack of Data, Inaccurate Data, and Limited Distribution. To effectively target a specific group, traditional marketing relies heavily on demographic information. The target audience may not be accurately represented even with demographic data. A campaign aimed at millennial women might overlook some individuals who don't fit the typical demographic. The target audience may not be reached effectively through traditional marketing channels. The desired audience might not see a print ad in a local newspaper if they don't read it. Traditional marketing doesn't provide immediate feedback on a campaign's success, making it difficult to see a return on investment or adapt messaging immediately. There can be a lot of wasted resources and opportunities due to limited targeting. In conjunction with AI technology, traditional marketing can be more targeted to the right audience by using a combination of traditional marketing and AI technology.

## 13.2. Time-Consuming

Traditional marketing can be time-consuming due to its inherent processes. Planning and executing campaigns, printing and distributing materials, and meeting with clients or customers are some of the reasons why.Multiple planning stages are involved in traditional marketing campaigns, including market research, strategy development, and creative development. Each stage requires meticulous planning and careful consideration, which can prolong the process. The quantity and mode of distribution of traditional marketing materials can affect the time it takes to create and distribute them. Producing promotional flyers, for instance, requires designing content, finalizing the layout, printing, and manually distributing or mailing them. Relationships are built through face-to-face interactions, which require meetings and presentations. Traditional marketing efforts can also be time-consuming to track and measure, requiring manual tracking of customer engagement, sales, and awareness campaigns. Due to manual processes, human interactions, and slow measurement methods, traditional marketing is time-consuming. Providing a personalized touch to customers is more important than ever as the world becomes more tech-savvy and fast-paced.

## 13.3. Inability to Measure Success Accurately

It can be difficult to measure the effectiveness of a traditional marketing campaign. Unlike online marketing, the effectiveness of an advertisement cannot be measured. It is difficult to track campaign results, rely on unreliable data, and target customers with high precision: Polls and surveys are often used to measure success, but they can be unreliable and inconsistent. In some cases, customer responses are delayed, erratic, or incomplete, making it difficult to get a clear picture of what is going on.

With traditional marketing, it's difficult to know what's working and what's not because tracking isn't as precise as it is with digital marketing. Traditional marketing can result in less accurate targeting, complicating the evaluation of specific campaigns. Challenges in Measuring Customer Engagement: It's harder to measure engagement levels with traditional marketing methods compared to digital ones. Traditional marketing has an unclear impact because it is difficult to track campaign results, customer targeting is less precise, and customer engagement can be difficult to track. Traditional marketing efforts are hard to measure, making planning for the future difficult.

## 15. Conclusion

During the past few years, Artificial Intelligence (AI) has significantly changed the landscape of food marketing. Advertising techniques that were once the backbone of marketing strategies are now being supplemented and often replaced by AI-based methods. Consumer marketing messages are changing as a result of this transformation. Using machine learning, natural language processing, and predictive analytics, artificial intelligence enables businesses to personalize recommendations, forecast consumer behavior, and optimize their marketing efforts. TV, radio, print, and outdoor advertising have been the main methods of traditional food marketing. Although these methods are effective in building brand awareness and reaching large audiences, modern consumers often expect personalized engagement. Marketers can now craft more targeted, relevant, and engaging campaigns thanks to digital media and artificial intelligence (AI). AI is used extensively in food marketing to develop customized recommendation systems. Various sources of data, including purchase histories, browsing behavior, and social media interactions, are used in these systems to deliver highly tailored product recommendations. Traditional advertising techniques that rely on demographic information cannot analyze individual preferences and predict future behaviors as accurately as AI-driven recommendation systems. By personalizing the experience for consumers, we are able to increase customer satisfaction, enhance the consumer experience, and increase conversion rates. Artificial intelligence is also making a significant impact on predictive analytics. The use of historical data can help marketers anticipate consumer needs and behaviors by predicting future outcomes.

- Personalization and Consumer Engagement

Food marketing benefits greatly from AI's ability to develop personalized recommendation systems. These systems provide highly personalized product recommendations based on information from multiple sources, such as purchase histories, browsing patterns, and social media interactions. Artificial intelligence can analyze individual preferences and forecast future behaviors more accurately than traditional advertising methods, which rely on broad demographic data. As a result, customer satisfaction and consumer experience are improved, as well as conversion rates are increased.

- Predictive Analytics and Market Trends

In food marketing, AI has also transformed predictive analytics. Marketers can predict consumer behavior and needs using AI algorithms by examining historical data. Food marketers can use this capability to identify trends, optimize inventory management, and design effective marketing campaigns. Marketing professionals can use AI to predict seasonal trends and consumer preferences, allowing them to promote the right products to the right audiences at the right time. Marketing efforts are more effective when they are timely and relevant, due to this strategic foresight.

- Efficiency and Cost-Effectiveness

Cost-effectiveness and efficiency are significant advantages of AI-driven strategies. The planning, execution, and monitoring of traditional advertising campaigns require significant time and resources. A number of these processes can be automated with AI tools, reducing the need for manual intervention. Programmatic advertising platforms, for example, automate real-time ad buying and placement based on consumer behavior and engagement metrics. By automating the advertising process, budget allocation is optimized, and return on investment is maximized.

- Dynamic and Responsive Marketing

By integrating AI into food marketing, consumers can be engaged in a more dynamic and responsive manner. There are often fixed schedules and content in traditional advertising methods, making them unable to adapt to changing consumer preferences and market conditions. Marketing messages can, however, be continuously adjusted by AI-driven strategies based on consumer interaction. By adapting to emerging trends, gathering consumer feedback, and refining campaigns, marketers can improve their results. AI-powered chatbots, for example, can provide personalized recommendations, answer questions, and collect data to enhance future marketing efforts.

- Data Insights and Consumer Behavior

As a result of AI technologies, marketers have access to large volumes of consumer data, allowing them to gain a deeper understanding of how consumers behave. In traditional advertising methods, consumers' preferences are often identified by aggregated data and general market research. Analyzing data using AI, however, reveals patterns and correlations that are difficult to discover with other types of analytics. Marketing professionals can use these insights to create more accurate customer profiles, segment audiences more effectively, and design highly targeted campaigns with these insights.

- Challenges and Ethical Considerations

Artificial intelligence-driven strategies offer many advantages, but they also pose some challenges and ethical concerns. Data privacy and security are two of the most pressing concerns. Large amounts of consumer data are collected, stored, and used in food marketing, raising questions about their use. It is imperative that marketers adhere to data protection regulations and adopt transparent practices in order to build consumer trust. As a result of the volume of targeted marketing messages, consumers may feel overwhelmed and engage less. In order to maintain consumer trust, personalization and privacy must be balanced.

- Investment and Expertise

Technology and expertise are required to implement AI in food marketing. The initial setup costs of AI-driven strategies can be high, but they can save money and improve efficiency in the long run. A marketer's ability to stay competitive depends on investing in advanced analytics tools, hiring skilled data scientists, and updating their technology on a regular basis. As AI algorithms become more complex, marketing professionals will also need ongoing training and education to fully exploit their potential.

## 15.1. *Comparative Analysis and Future Research*

Traditional advertising techniques and AI-driven strategies are compared in this study, highlighting their strengths and weaknesses. In the food marketing industry, artificial intelligence (AI) technologies are revolutionizing engagement, satisfaction, and conversion by enhancing consumer engagement, satisfaction, and conversion. By automating and analyzing data in real-time, AI-driven marketing strategies are more efficient and cost-effective than traditional marketing methods. Investing in AI technologies, training staff, and continuously updating marketing approaches are all necessary steps for food marketers to stay competitive in the digital landscape. In the future, the long-term effects of AI-driven marketing on consumer behavior should be explored, as well as potential challenges such as over-personalization. AI also offers ongoing opportunities to refine and improve marketing strategies due to its evolving nature.

**References**

[1] N. J. Rowan, "The role of digital technologies in supporting and improving fishery and aquaculture across the supply chain – Quo Vadis?," *Aquac. Fish.*, vol. 8, no. 4, pp. 365–374, 2023, doi: 10.1016/j.aaf.2022.06.003.

[2] A. Ravikumar, *The role of machine learning and computer vision in the agri-food industry*. 2022. doi: 10.4018/978-1-6684-5141-0.ch014.

[3] C. Yaiprasert and A. N. Hidayanto, "AI-driven ensemble three machine learning to enhance digital marketing strategies in the food delivery business," *Intell. Syst. with Appl.*, vol. 18, 2023, doi: 10.1016/j.iswa.2023.200235.

[4] M. Philp, J. Jacobson, and E. Pancer, "Predicting social media engagement with computer vision: An examination of food marketing on Instagram," *J. Bus. Res.*, vol. 149, pp. 736 – 747, 2022, doi: 10.1016/j.jbusres.2022.05.078.

[5] Z. Liu, "Unlocking Consumer Choices in the Digital Economy: Exploring Factors Influencing Online and Offline Purchases in the Emerging Pet Food Market," *J. Knowl. Econ.*, 2023, doi: 10.1007/s13132-023-01490-8.

[6] G. Califano and C. Spence, "Assessing the visual appeal of real/AI-generated food images," *Food Qual. Prefer.*, vol. 116, 2024, doi: 10.1016/j.foodqual.2024.105149.

[7]     D. Chiras, M. Stamatopoulou, N. Paraskevis, S. Moustakidis, I. Tzimitra-Kalogianni, and C. Kokkotis, "Explainable Machine Learning Models for Identification of Food-Related Lifestyle Factors in Chicken Meat Consumption Case in Northern Greece," *BiomedInformatics*, vol. 3, no. 3, pp. 817 – 828, 2023, doi: 10.3390/biomedinformatics3030051.

[8]     R. Brooks *et al.*, "Use of artificial intelligence to enable dark nudges by transnational food and beverage companies: analysis of company documents," *Public Health Nutr.*, vol. 25, no. 5, pp. 1291 – 1299, 2022, doi: 10.1017/S1368980022000490.

[9]     P. Clark, J. Kim, and Y. Aphinyanaphongs, "Marketing and US Food and Drug Administration Clearance of Artificial Intelligence and Machine Learning Enabled Software in and as Medical Devices: A Systematic Review," *JAMA Netw. Open*, vol. 6, no. 7, p. E2321792, 2023, doi: 10.1001/jamanetworkopen.2023.21792.

[10]    M. S. Wetherill *et al.*, "Food choice considerations among American Indians living in rural Oklahoma: The THRIVE study," *Appetite*, vol. 128, pp. 14 – 20, 2018, doi: 10.1016/j.appet.2018.05.019.

[11]    E. Skawińska and R. I. Zalewski, "New Foods as a Factor in Enhancing Energy Security," *Energies*, vol. 17, no. 1, 2024, doi: 10.3390/en17010192.

[12]    V. V Zinchenko *et al.*, "Methodology for Conducting Post-Marketing Surveillance of Software as a Medical Device Based on Artificial Intelligence Technologies," *Sovrem. Tehnol. v Med.*, vol. 14, no. 5, pp. 15 – 25, 2022, doi: 10.17691/stm2022.14.5.02.

[13]    A. J. Roberts, T. W. Geary, E. E. Grings, R. C. Waterman, and M. D. Macneil, "Reproductive performance of heifers offered ad libitum or restricted access to feed for a one hundred forty-day period after weaning," *J. Anim. Sci.*, vol. 87, no. 9, pp. 3043 – 3052, 2009, doi: 10.2527/jas.2008-1476.

[14]    G. Bermúdez-González, E. M. Sánchez-Teba, M. D. Benítez-Márquez, and A. Montiel-Chamizo, "Generation z young people's perception of sexist female stereotypes about the product advertising in the food industry: Influence on their purchase intention," *Foods*, vol. 11, no. 1, 2022, doi: 10.3390/foods11010053.

[15]    A. Sestino, M. V. Rossi, L. Giraldi, and F. Faggioni, "Innovative food and sustainable

consumption behaviour: the role of communication focus and consumer-related characteristics in lab-grown meat (LGM) consumption," *Br. Food J.*, vol. 125, no. 8, pp. 2884 – 2901, 2023, doi: 10.1108/BFJ-09-2022-0751.

[16] A. C. L. Kuang, T. M. Lim, C. W. Tan, C. F. Ho, and N. A. Husaini, "AI Ads: Practicability of Text Generation for F&B Marketing," *J. Logist. Informatics Serv. Sci.*, vol. 11, no. 2, pp. 324 – 345, 2024, doi: 10.33168/JLISS.2024.0220.

[17] G. Camelo and M. Nogueira, "The ESG Menu: Integrating Sustainable Practices in the Portuguese Agri-Food Sector," *Sustain.*, vol. 16, no. 11, 2024, doi: 10.3390/su16114377.

[18] A. Hamdy, J. Zhang, and R. Eid, "A new proposed model for tourists' destination image formation: the moderate effect of tourists' experiences," *Kybernetes*, vol. 53, no. 4, pp. 1545 – 1566, 2024, doi: 10.1108/K-11-2022-1525.

[19] A. Dalgic and A. S. Demircioğlu Dalgıç, "Technological workforces of events: where and how to use them?," *Worldw. Hosp. Tour. Themes*, 2024, doi: 10.1108/WHATT-06-2024-0122.

[20] M. Jayanthi and D. Shanthi, "Predicting Crop Yield with AI—A Comparative Study of DL and ML Approaches," *Lect. Notes Networks Syst.*, vol. 840, pp. 337 – 348, 2024, doi: 10.1007/978-981-99-8451-0_29.

[21] R. S. Upendra, M. R. Ahmed, T. N. Kumar, S. R. Prithviraj, and A. S. Khan, "Indian Agricultural Sector Present and Post Pandemic Condition," *Indian J. Agric. Res.*, vol. 57, no. 1, pp. 1 – 7, 2023, doi: 10.18805/IJARe.A-5709.

[22] U. N. Dulhare and S. Gouse, "Automation of Rice Cultivation from Ploughing–Harvesting with Diseases, Pests and Weeds to Increase the Yield Using AI," *Lect. Notes Electr. Eng.*, vol. 828, pp. 505 – 513, 2022, doi: 10.1007/978-981-16-7985-8_51.

[23] Y. Li and W. Xu, "D-AdFeed: A diversity-aware utility-maximizing advertising framework for mobile users," *Comput. Networks*, vol. 190, 2021, doi: 10.1016/j.comnet.2021.107954.

[24] A. Ssematimba *et al.*, "Analysis of geographic location and pathways for influenza A virus infection of commercial upland game bird and conventional poultry farms in the United States of America," *BMC Vet. Res.*, vol. 15, no. 1, 2019, doi: 10.1186/s12917-019-1876-y.

[25] G. Richards, "The curatorial turn in tourism and hospitality," *Int. J. Contemp. Hosp.*

*Manag.*, vol. 36, no. 13, pp. 19 – 37, 2024, doi: 10.1108/IJCHM-06-2023-0905.

[26] J. Dutta, M. Patwardhan, P. Deshpande, S. Karande, and B. Rai, "Zero-shot transfer learned generic AI models for prediction of optimally ripe climacteric fruits," *Sci. Rep.*, vol. 13, no. 1, 2023, doi: 10.1038/s41598-023-34527-8.

[27] R. Dhillon and Q. Moncur, "Small-Scale Farming: A Review of Challenges and Potential Opportunities Offered by Technological Advancements," *Sustain.*, vol. 15, no. 21, 2023, doi: 10.3390/su152115478.

[28] S. Monaco and G. Sacchi, "Travelling the Metaverse: Potential Benefits and Main Challenges for Tourism Sectors and Research Applications," *Sustain.*, vol. 15, no. 4, 2023, doi: 10.3390/su15043348.

[29] Z. Xie, X. Liang, and C. Roberto, "Learning-based robotic grasping: A review," *Front. Robot. AI*, vol. 10, 2023, doi: 10.3389/frobt.2023.1038658.

[30] J. H. Woo, E. C. Kim, and S. M. Kim, "The current status of breakthrough devices designation in the United States and innovative medical devices designation in Korea for digital health software," *Expert Rev. Med. Devices*, vol. 19, no. 3, pp. 213–228, 2022, doi: 10.1080/17434440.2022.2051479.

[31] B. P. Singh, M. Chander, S. S. Pathade, and K. Pordhiya, "Livestock services delivery during COVID-19 lockdown: An appraisal of accessibility and constraints," *Indian J. Anim. Sci.*, vol. 91, no. 7, pp. 595 – 599, 2021, [Online]. Available: https://www.scopus.com/inward/record.uri?eid=2-s2.0-85115935623&partnerID=40&md5=1ea7562196c7649456b7e38f49268303

[32] C.-W. Wu and A. Monfort, "Role of artificial intelligence in marketing strategies and performance," *Psychol. Mark.*, vol. 40, no. 3, pp. 484 – 496, 2023, doi: 10.1002/mar.21737.

[33] M.-C. Chiu and K.-H. Chuang, "Applying transfer learning to achieve precision marketing in an omni-channel system–a case study of a sharing kitchen platform," *Int. J. Prod. Res.*, vol. 59, no. 24, pp. 7594 – 7609, 2021, doi: 10.1080/00207543.2020.1868595.

[34] H. Coope, L. Parviainen, M. Withe, J. Porter, and R. J. Ross, "Hydrocortisone granules in capsules for opening (Alkindi) as replacement therapy in pediatric patients with adrenal insufficiency," *Expert Opin. Orphan Drugs*, vol. 9, no. 3, pp. 67 – 76, 2021, doi:

10.1080/21678707.2021.1903871.

[35] X. Li, H. Liang, and Z. Liu, "Health Claims Unpacked: A toolkit to Enhance the Communication of Health Claims for Food," in *International Conference on Information and Knowledge Management, Proceedings*, 2021, pp. 4744 – 4748. doi: 10.1145/3459637.3481984.

[36] S. M. R. Rahman, M. N. Islam, M. D. Harun-ur-Rashid, M. S. R. Siddiki, and M. A. Islam, "Dairy buffalo production under intensive system in semi arid area of Bangladesh," *Buffalo Bull.*, vol. 38, no. 1, pp. 83 – 98, 2019, [Online]. Available: https://www.scopus.com/inward/record.uri?eid=2-s2.0-85065253990&partnerID=40&md5=ba8b09b2044d51992963eaedef0183a6

[37] M. Schirra, P. Cabras, A. Angioni, G. D'Hallewin, and M. Pala, "Residue uptake and storage responses of Tarocco blood oranges after preharvest thiabendazole spray and postharvest heat treatment," *J. Agric. Food Chem.*, vol. 50, no. 8, pp. 2293 – 2296, 2002, doi: 10.1021/jf0114583.

[38] K. S. Yee, T. E. Carpenter, S. Mize, and C. J. Cardona, "The live bird market system and low-pathogenic avian influenza prevention in Southern California," *Avian Dis.*, vol. 52, no. 2, pp. 348 – 352, 2008, doi: 10.1637/8138-101207-Reg.1.

[39] H. Ding *et al.*, "The Application of Artificial Intelligence and Big Data in the Food Industry," *Foods*, vol. 12, no. 24, 2023, doi: 10.3390/foods12244511.

[40] L. Nordmark *et al.*, "Launch of IoT and artificial intelligence to increase the competitiveness in Swedish apple and grapevine production," *Acta Hortic.*, vol. 1314, pp. 235–240, 2021, doi: 10.17660/ActaHortic.2021.1314.30.

[41] S. Sharma, V. K. Gahlawat, K. Rahul, R. S. Mor, and M. Malik, "Sustainable Innovations in the Food Industry through Artificial Intelligence and Big Data Analytics," *Logistics*, vol. 5, no. 4, 2021, doi: 10.3390/logistics5040066.

[42] N. Zahradka, J. Pugmire, J. L. Taylor, A. Wolfberg, and M. Wilkes, "Deployment of an End-to-End Remote, Digitalized Clinical Study Protocol in COVID-19: Process Evaluation," *JMIR Form. Res.*, vol. 6, no. 7, 2022, doi: 10.2196/37832.

[43] E. Willis, K. Friedel, M. Heisten, M. Pickett, and A. Bhowmick, "Communicating Health

Literacy on Prescription Medications on Social Media: In-depth Interviews With 'Patient Influencers,'" *J. Med. Internet Res.*, vol. 25, 2023, doi: 10.2196/41867.

[44] P. D. Gaspar, V. N. G. J. Soares, J. M. L. P. Caldeira, L. P. Andrade, and C. D. Soares, "Technological modernization and innovation of traditional agri-food companies based on ICT solutions—The Portuguese case study," *J. Food Process. Preserv.*, vol. 46, no. 8, 2022, doi: 10.1111/jfpp.14271.

[45] K. E. Axenov, O. V Kraskovskaya, and F. M. Renni, "SPATIAL ORGANIZATION OF THE NEW FORMS OF E-GROCERY AND READY-MADE FOOD TRADE IN A LARGE RUSSIAN CITY," *Balt. Reg.*, vol. 14, no. 3, pp. 28 – 48, 2022, doi: 10.5922/2079-8555-2022-3-2.

[46] B. C. Mazzolani *et al.*, "Disordered Eating Attitudes and Food Choice Motives Among Individuals Who Follow a Vegan Diet in Brazil," *JAMA Netw. Open*, vol. 6, no. 6, p. E2321065, 2023, doi: 10.1001/jamanetworkopen.2023.21065.

[47] M. Amerzadeh, A. Takian, H. Pouraram, A. A. Sari, and A. Ostovar, "Policy analysis of socio-cultural determinants of salt, sugar and fat consumption in Iran," *BMC Nutr.*, vol. 8, no. 1, 2022, doi: 10.1186/s40795-022-00518-7.

[48] V. Russo *et al.*, "The Role of the Emotional Sequence in the Communication of the Territorial Cheeses: A Neuromarketing Approach," *Foods*, vol. 11, no. 15, 2022, doi: 10.3390/foods11152349.

[49] D. Dikmen *et al.*, "Cross-Sectional Evaluation of Food Items Preferred by Adolescents under the Influence of Television Advertisements," *J. Res. Health Sci.*, vol. 22, no. 1, 2022, doi: 10.34172/jrhs.2022.74.

[50] F. K. K. Putra, M. K. Putra, and S. Novianti, "Taste of asean: traditional food images from Southeast Asian countries," *J. Ethn. Foods*, vol. 10, no. 1, 2023, doi: 10.1186/s42779-023-00189-0.

[51] N. Ahmad, L. S. B. H. Shamsu, and M. D. I. Ariffin, "Halal Meat, Food Fraud, and Consumer Protection: A Comparison of Islamic, European and Malaysian Perspectives," *Manchester J. Transnatl. Islam. Law Pract.*, vol. 19, no. 2, pp. 80 – 98, 2023, [Online]. Available: https://www.scopus.com/inward/record.uri?eid=2-s2.0-

85166416911&partnerID=40&md5=357b398cf0a2fbc2d425224ad52fbec6

[52] L. Czaplicki, K. Marynak, D. Kelley, M. B. Moran, S. Trigger, and R. D. Kennedy, "Presence of Nicotine Warning Statement on US Electronic Nicotine Delivery Systems (ENDS) Advertisements 6 Months Before and After the August 10, 2018 Effective Date," *Nicotine Tob. Res.*, vol. 24, no. 11, pp. 1720 – 1726, 2022, doi: 10.1093/ntr/ntac104.

[53] A. B. Jibril and D. E. Adzovie, "Understanding the moderating role of E-WoM and traditional media advertisement toward fast-food joint selection: a uses and gratifications theory," *J. Foodserv. Bus. Res.*, vol. 27, no. 1, pp. 1 – 25, 2022, doi: 10.1080/15378020.2022.2070450.

[54] N. T. Giannakopoulos, M. C. Terzi, D. P. Sakas, N. Kanellos, K. S. Toudas, and S. P. Migkos, "Agroeconomic Indexes and Big Data: Digital Marketing Analytics Implications for Enhanced Decision Making with Artificial Intelligence-Based Modeling," *Inf.*, vol. 15, no. 2, 2024, doi: 10.3390/info15020067.

[55] S. Hackfort, S. Marquis, and K. Bronson, "Harvesting value: Corporate strategies of data assetization in agriculture and their socio-ecological implications," *Big Data Soc.*, vol. 11, no. 1, 2024, doi: 10.1177/20539517241234279.

[56] R. Singh, J. Kumar, and A. Nayak, "AGROY: creating value through smart farming," *Emerald Emerg. Mark. Case Stud.*, vol. 9, no. 3, pp. 1 – 31, 2019, doi: 10.1108/EEMCS-10-2018-0214.

[57] M. N. I. Sarker, M. S. Islam, H. Murmu, and E. Rozario, "Role of big data on digital farming," *Int. J. Sci. Technol. Res.*, vol. 9, no. 4, pp. 1222 – 1225, 2020, [Online]. Available: https://www.scopus.com/inward/record.uri?eid=2-s2.0-85083496987&partnerID=40&md5=2f2f612b053729c968940e9efc8f94f9

[58] P. Jayashankar, W. J. Johnston, S. Nilakanta, and R. Burres, "Co-creation of value-in-use through big data technology- a B2B agricultural perspective," *J. Bus. Ind. Mark.*, vol. 35, no. 3, pp. 508 – 523, 2020, doi: 10.1108/JBIM-12-2018-0411.

[59] B. Mao and H. Tian, "Business model based on the synergistic drive of flexible supply chain and digital marketing," *Int. J. Inf. Commun. Technol.*, vol. 24, no. 8, pp. 1 – 19, 2024, doi: 10.1504/IJICT.2024.139868.

[60] T. Yigitcanlar *et al.*, "Artificial intelligence technologies and related urban planning and development concepts: How are they perceived and utilized in Australia?," *J. Open Innov. Technol. Mark. Complex.*, vol. 6, no. 4, pp. 1 – 21, 2020, doi: 10.3390/joitmc6040187.